\documentclass[11pt,a4paper]{article}

\usepackage[paperwidth=26cm]{geometry}
\usepackage[T1]{fontenc}
\usepackage[latin1]{inputenc}
\usepackage{graphicx}
\usepackage{amsmath}
\usepackage{amssymb}

\newcommand{\bl}{\boldsymbol}

\newcommand{\lef}{\left(\,}
\newcommand{\rig}{\,\right)}
\newcommand{\ph}{\phantom}

\newcommand{\eq}{\,=\,}
\newcommand{\ma}{\,+\,}
\newcommand{\me}{\,-\,}

\begin{document}

\title{\textbf{Conformally Invariant Spinorial Equations in Six Dimensions }}

\author{\textbf{Carlos Batista}\\
\small{Departamento de F\'{\i}sica}\\
\small{Universidade Federal de Pernambuco}\\
\small{50670-901 Recife-PE, Brazil}\\
\small{carlosbatistas@df.ufpe.br}}
\date{\today}




\maketitle

\begin{abstract}
This work deals with the conformal transformations in six-dimensional spinorial formalism. Several conformally invariant equations are obtained and their geometrical interpretation are worked out. Finally, the integrability conditions for some of these equations are established. Moreover, in the course of the article, some useful identities involving the curvature of the spinorial connection are attained and a digression about harmonic forms and more general massless fields is made.
\textsl{(Keywords: Spinors, Conformal transformations, Twistors, Massless fields, Six dimensions, Integrability Conditions)}
\end{abstract}

\section{Introduction}

Although it is possible to obtain general results handling spinors without specifying the space dimension, in the spinorial formalism each dimension has its own peculiarities. The easiest way to identify and take advantage of these  particularities is by means of the index notation. In such approach, a spinor is represented by an object possessing one index running from 1 to $2^{[n/2]}$, where $n$ denotes the dimension of the space and $[n/2]$ stands for the integer part of $n/2$. Particularly, in even dimensions, a general spinor can be decomposed as the sum of two Weyl spinors of opposite chiralities, each Weyl spinor possessing $2^{\frac{n-2}{2}}$ components. For instance, in four dimensions the Weyl spinors are objects with indices ranging from 1 to 2. This two-component formalism, introduced by Roger Penrose, was the basis for all the great achievements of the spinorial  calculus in four-dimensional general relativity \cite{Penrose-Spinors-in-GR,Peeling-Penrose,Penrose-Books}. For example, the two-component spinorial formalism played a prominent role in understanding the Petrov classification, the Goldberg-Sachs theorem and the asymptotic behaviour of massless fields in asymptotically flat spacetimes. Even though these topics can be grasped without spinors, it is much more enlightening to address them using the spinorial formalism in its index notation. Moreover, the two-component approach  brought various advances for the study of four-dimensional geometrical structures, as exemplifies the twistor theory. With such applications in mind, the aim of the present article is to investigate and obtain conformally invariant equations in six-dimensional spaces using the spinorial index formalism. Alternative ways of attacking the problem of finding conformally invariant operators can be found in \cite{Fischmann,Faci13,Tractor} and references therein.


The applications of conformal invariance in physics dates back to 1909, when H. Bateman and E. Cunningham noticed that Maxwell's equations were invariant under conformal transformations \cite{Bateman,Cunningham-1910}. A decade later, Hermann Weyl took another huge step when he introduced the concept of gauge freedom, while he  unsuccessfully  attempted  to bind gravitation and electromagnetism in a unique geometrical theory \cite{Weyl18}. Since that time, the notion of conformal invariance has acquired an increasingly relevance in physical theories. Classically, massless particles move along null directions, which, in turn, are invariant under conformal transformations. Thus, it is reasonable to expect that massless field equations are conformally invariant. However, this guess turn out to be erroneous, as there are plenty of massless equations of motion that are not invariant under conformal transformations, as exemplified along this article. Therefore, those massless theories that are conformally invariant are endowed with quite special mathematical and physical properties \cite{DiFrancesco}. In particular, over the past decade much attention has been drawn to the AdS/CFT correspondence, which provides access to non-perturbative aspects of quantum gravity and enable the treatment of strongly coupled conformal field theories, with relevance for condensed matter and nuclear physics \cite{AdS-CFT-Maldacena,AdS-CFT-2}. Scale invariance also plays a central role in statistical physics and quantum field theory, as this symmetry is present in the fixed points of the renormalization group \cite{RenormalizationGroupReviews}. Although the scale symmetry is less stringent than the conformal symmetry, it turns out that almost all physical applications of the former comes accompanied by the latter symmetry \cite{Nakayama-Scale-vs-Conf}. For an historical account on physical applications of the conformal transformations, the reader is referred to \cite{Kastrup}.

The conformal group of the space $\mathbb{R}^{p,q}$ is given by $SO(p+1,q+1)$. Thus, the conformal group in four dimensions is the rotation group in a six-dimensional space, which gives one important physical motivation for investigating spaces of six dimensions. Indeed, this route for studying four-dimensional conformal field theory has already been taken  \cite{Dirac-Conf6D,Weinberg-Conf6D,Drew-72,Barut-Conf6D}. Still, spaces of six dimensions are also interesting in themselves, as they can be used to model higher-dimensional theories of (super-)gravity \cite{Review-ModifiedGravity}. In addition, it has been constructed a correspondence between a six-dimensional conformal field theory and a gravitational theory in $AdS_7\times S^4$ \cite{AdS-CFT6}. Furthermore, the computation of scattering amplitudes of massless fields by means of the spinor-helicity formalism have also been put forward in six dimensions \cite{Spinor-helicity6D-1,Spinor-helicity6D-2}. Moreover, the six-dimensional twistor theory and the Penrose transform have recently been considered in \cite{Mason:2011nw,Saemann:2011nb}.

The structure of this article goes as follows. In Sec. \ref{Sec.Spinor6D}, the index formalism for spinors in six dimensions is introduced and the notation used throughout the article is established. Then, the consequences of a conformal transformation on the spinorial connection and its curvature are displayed in Sec. \ref{Sec.ConfTransf}. The latter results are then used to investigate the conformal properties of some massless fields in Sec. \ref{Sec.MasslessFields}. More precisely, we consider how the harmonic condition on differential forms is represented in the spinorial formalism and how this condition is affected by a conformal transformation. Inspired by a few well-known conformally invariant equations for fields of spin $1/2$, in Sec. \ref{Sec.ConformallyIvariantEq} we present a series of conformally invariant equations for massless fields of higher spin. We further provide a geometrical interpretation for these fields. Moreover, in Sec. \ref{Sec.IntegrabilityCond}, we obtain the integrability conditions for some of these equations. Finally, in Sec. \ref{Sec.Conclusions}, the concluding remarks are presented.



\section{Spinorial Formalism in Six Dimensions}\label{Sec.Spinor6D}

The aim of this section is to briefly review the index approach for the spinorial formalism in six dimensions as well as to set the notation adopted throughout the article. A thorough introduction to this subject can be found in \cite{Bat-Spin6D,Bat-Book}.

The group $SPin(\mathbb{R}^{p,q})$ is the double covering of $SO(\mathbb{R}^{p,q})$, and the space of spinors is the vectorial space in which acts the  faithful representation of the group $SPin(\mathbb{R}^{p,q})$ with lower dimension. For instance, $SPin(\mathbb{R}^{3,1})\sim SL(2,\mathbb{C})$ is the double covering of the Lorentz group $SO(3,1)$. This is the theoretical basis behind the 2-component spinorial formalism introduced by Roger Penrose in four-dimensional spacetimes. These spinors with two components are the objects in which a 2-dimensional representation of the group $SL(2,\mathbb{C})$ acts. There are two independent 2-dimensional representations of $SL(2,\mathbb{C})$, the traditional one that comes from the very definition of this group and the complex conjugate of the latter, which cannot be obtained from the first representation by a similarity transformation. The objects in which the first of these representations act are the Weyl spinors of positive chirality, while objects transforming according to the second representation are Weyl spinors of negative chirality. In the same fashion, in six-dimensional Euclidean spaces, we have that $SPin(\mathbb{R}^{6})\sim SU(4)$. Thus, in a six-dimensional space of Euclidean signature, spinors are objects that transform in the 4-dimensional representations of $SU(4)$, namely objects $\chi^A$ and $\lambda_A$ such that the action of  $\bl{U}\in SU(4)$ is given by
\begin{equation}\label{SU(4)-fund.}
  \chi^{A}\,\stackrel{\bl{U}}{\longrightarrow} \,U^A_{\phantom{A}B} \,\chi^{B}\;\;\;\;\; ,\, \;\;\;\;\; \lambda_{A}\,\stackrel{\bl{U}}{\longrightarrow}\, U^{-1\,B}_{\phantom{-1\,B}A}\, \lambda_{B}\,\,.
\end{equation}
Where the capital indices $A,B,\ldots$ range from 1 to 4 and $U^A_{\phantom{A}B}$ is a unitary matrix of unit determinant, with $U^{-1\,B}_{\phantom{-1\,B}A}$ being its inverse. In particular, it follows that the scalar $\chi^{A}\lambda_{A}$ is invariant by the action of  $SU(4)$. The spinors $\chi^{A}$ are said to be Weyl spinors of positive chirality, while $\lambda_{A}$ are Weyl spinors of negative chirality. A Dirac spinor is then given by a pair $(\chi^{A},\lambda_{B})$.

The representations carried by $\chi^{A}$ and $\lambda_{A}$ are the only independent 4-dimensional faithful representations of the group $SU(4)$. However, in six-dimensional spaces of non-Euclidean signatures there exist two more independent representations. A convenient way of handling the case of arbitrary signature is by considering the complex case and, when necessary, imposing reality conditions in the vector space in order to choose a signature, in the spirit of \cite{Bat-Book,Bat-art4}. Therefore, let us consider the spinors of the vector space $\mathbb{R}^{6}\otimes \mathbb{C}\sim \mathbb{C}^{6}$. Inasmuch as $SPin(\mathbb{C}^{6})\sim SL(4,\mathbb{C})$, and since $SL(4,\mathbb{C})$ admit four independent 4-dimensional representations, we can have four types of spinors in the complex case. The actions of an element $\bl{S}\in SL(4,\mathbb{C})$ in these representations are given by
$$  \chi^{A}\stackrel{\bl{S}}{\longrightarrow} S^A_{\ph{A}B} \,\chi^{B}\;\;\;\; , \;\;\;\;
          \lambda_{A}\stackrel{\bl{S}}{\longrightarrow} S^{-1\,B}_{\phantom{-1\,B}A} \,\lambda_{B} \;\;\;\; , \;\;\;\; \lambda_{\dot{A}} \stackrel{\bl{S}}{\longrightarrow} \overline{S}_{\dot{A}}^{\phantom{A}\dot{B}}\, \lambda_{\dot{B}} \;\;\;\; , \;\;\;\;
        \chi^{\dot{A}}\stackrel{\bl{S}}{\longrightarrow} \overline{S}^{\,-1\phantom{B}\dot{A}}_{\phantom{-1}\dot{B}} \,\chi^{\dot{B}}\,.  $$
Where $S^A_{\ph{A}B}$ is a $4\times 4$ complex matrix of unit determinant, with $S^{-1\,B}_{\phantom{-1\,B}A}$,  $\overline{S}_{\dot{A}}^{\phantom{A}\dot{B}}$ and $\overline{S}^{\,-1\phantom{B}\dot{A}}_{\phantom{-1}\dot{B}}$ being its inverse, its complex conjugate and the inverse of its complex conjugate respectively. The dot over the spinorial index serves to recall that we are dealing with a complex conjugated representation.  The reason for the representations $\chi^{\dot{A}}$ and  $\lambda_{\dot{B}}$ to be unnecessary in the Euclidean signature is that in the group $SU(4)$ the complex conjugate of the 4-dimensional matrix representation is just the transpose of the inverse. In the general case, dotted indices can be transformed into undotted indices just by means of the introduction of a charge conjugation operator. In what follows we shall deal just with undotted indices, but completely analogous results follow for spinors with dotted indices.

Since $SPin(\mathbb{C}^{6})$ is a double covering for the group $SO(\mathbb{C}^{6})$, it follows that one can use the spinor to generate the vectorial and tensorial representations of the group $SO(\mathbb{C}^{6})$, as illustrated in \cite{Bat-Spin6D}. For instance, a vector in $\mathbb{C}^6$ possess six components, so that its spinorial representation must also have six components. Indeed, the spinorial equivalent of a vector in $\mathbb{C}^6$ is an object of the form $V^{AB}$ which is antisymmetric in its indices. Actually, in arbitrary dimensions, any totally skew-symmetric tensor is represented by an object with two spinorial indices, which stems from the fact that the space of differential forms is spanned by the quadratic objects $\langle \psi, \gamma_{a_1\cdots a_p} \varphi\rangle$, where $\psi$ and $\varphi$ are Dirac spinors,  $\gamma_{a_1\cdots a_p}$ is the skew-symmetric product of the Dirac matrices $\gamma_a$, and $\langle\,,\rangle$ is an inner product in the space of spinors that is invariant by the connected component of $SPin(\mathbb{R}^{p,q})$. Thus, for instance, in six dimensions a bivector $B_{ab} = -B_{ba}$ is represented by an object of the form $B^A_{\ph{A}B}$ that is trace-less. The latter object is in a 15-dimensional representation of the group $SU(4)$, in accordance with the fact that the space of bivectors over $\mathbb{C}^{6}$ has 15 dimensions.  Besides the vectors and bivectors, table \ref{Table spinors equivalent} summarizes the spinorial representations of some other important tensors in $\mathbb{C}^{6}$.
In the mentioned table, the indices $a,b,c,\cdots$ represent vectorial indices and, as usual, indices enclosed inside round brackets must be symmetrized, while indices inside square brackets are antisymmetrized. Such table have been provided here for the convenience of the reader, but an analogous table, as well as a careful explanation of its assertions, is available in Ref. \cite{Bat-Spin6D}.\\

\begin{table}[!htbp]
\begin{center}
\begin{tabular}{c c c}
  $SO(\mathbb{C}^{6})$ Form \quad\quad\quad\;& Spinorial Form & Algebraic Symmetries   \\ \hline\hline
  $V^a$ & $V^{AB}$ & $V^{AB}=V^{[AB]}$ \\
  $S_{ab}\,=\,S_{(ab)}\,,\,S^a_{\ph{a}a}=0$ &$ S^{AB}_{\phantom{AB}CD}$ &\quad\quad  $S^{AB}_{\phantom{AB}CD}=S^{[AB]}_{\phantom{AB}[CD]}, \, S^{AB}_{\phantom{AB}CB}=0$ \\
  $g_{ab}$ & $\frac{1}{2}\,\varepsilon_{ABCD}$ & $\varepsilon_{ABCD}\,=\, \varepsilon_{[ABCD]}$ \\
  $B_{ab}\,=\,B_{[ab]}$ & $B^A_{\phantom{A}B}$ & $B^A_{\phantom{A}A}=0$ \\
  $T_{abc}\,=\,T_{[abc]}$  & $(T^{AB},T_{AB})$& $T^{AB}=T^{(AB)}, T_{AB}=T_{(AB)}$  \\
  $C_{abcd}$ & $C^{AB}_{\phantom{AB}CD}$ & \quad\quad$C^{AB}_{\phantom{AB}CD}=C^{(AB)}_{\phantom{AB}(CD)},  C^{AB}_{\phantom{AB}CB} = 0$ \\
  \hline\hline
\end{tabular}
\caption{The first column gives the tensor structure in the space $\mathbb{C}^6$. The tensor $g_{ab}$ represents the metric of the vector space and $C_{abcd}$ is an object with the same algebraic symmetries of a Weyl tensor, namely $C_{abcd}=C_{[cd][ab]}$ and $C^a_{\;\,\,bad}=0$. The second column gives the spinorial equivalent for each tensor of the first column. The third column gives the symmetries that must be satisfied by these spinorial objects.  } \label{Table spinors equivalent}
\end{center}
\end{table}

Since a vector index is represented by a pair of skew-symmetric spinorial indices, $a\sim AB$, the action of lowering or raising a vectorial index by means of the metric $g_{ab}$ and its inverse $g^{ab}$ is represented in the spinorial language by lowering or raising a pair of anti-symmetric indices:
\begin{equation}\label{spinorialmetric}
  V_a\eq  g_{ab}\,V^b   \quad \sim \quad  V_{AB}\,=\,\frac{1}{2} \, \varepsilon_{ABCD}\,V^{CD} \quad\quad \textrm{and } \quad\quad V^a\eq  g^{ab}\,V_b   \quad \sim \quad  V^{AB}\,=\,\frac{1}{2} \, \varepsilon^{ABCD}\,V_{CD} \,.
\end{equation}
Where $\varepsilon_{ABCD}$ is the unique totally antisymmetric object such that $\varepsilon_{1234}=1$. But, it is worth pointing out that, differently from the well-known four-dimensional case, there is no natural way of raising or lowering a single spinorial index in six dimensions. Using the fact that the 4-dimensional representation of  $SL(4,\mathbb{C})$ is formed by matrices of unit determinant, it is simple matter to verify that the objects $\varepsilon_{ABCD}$  and $\varepsilon^{ABCD}$ are invariant by the group $SPin(\mathbb{C}^{6})$, just as the metric $g_{ab}$ is invariant by the action of the group $SO(\mathbb{C}^{6})$. Since a vectorial index is represented by an anti-symmetric pair of spinorial indices, one could wonder why a bivector $B_{ab}=B_{[ab]}$ is represented by $B^A_{\ph{A}B}$ instead of an object of the form $\mathfrak{B}_{AB\,CD}= \mathfrak{B}_{[AB]\,[CD]} = - \mathfrak{B}_{CD\,AB}$, which clearly has the equivalent of two anti-symmetric vectorial indices, as a bivector should have. The answer is that both objects are in the same representation of $SL(4,\mathbb{C})$. Indeed, the invariant object $\varepsilon^{ABCD}$ provides a one-to-one map between them,
$$ B^A_{\ph{A}B} \eq \frac{1}{4}\,\varepsilon^{AEDC}\,\mathfrak{B}_{DC\,EB}  \,.  $$
A detailed description of similar relations for the other tensors in table \ref{Table spinors equivalent} is available in Ref. \cite{Bat-Spin6D}.

Now, let us move from fixed six-dimensional vector spaces to tangent spaces of six-dimensional manifolds endowed with a metric $g_{ab}$ and equipped with the Levi-Civita connection, which is denoted here by $\nabla_a$. In order for a manifold to admit a spinor bundle it must satisfy some topological requirements, which will be assumed to hold throughout this article \cite{SpinStructure,BennTucker}. It turns out that the Levi-Civita connection can be extended to the spinor bundle and the extension is unique apart from a $U(1)$ gauge symmetry \cite{ConnectionSpin,Bat-PureSubspaces}. This gauge symmetry is fixed if we impose that such connection satisfies the Leibniz rule with respect to the natural inner product in the spinor bundle. As usual, this gauge fixing condition will be assumed here. This spinorial connection will also be denoted by $\nabla_a$, or $\nabla_{AB}$ in the spinorial representation. In particular, the curvature of such connection is given by
\begin{equation}\label{SpinorCurvature1}
 (\nabla_a\,\nabla_b \me \nabla_b\,\nabla_a)\, \chi^C \eq \mathfrak{R}_{ab\phantom{C}D}^{\ph{ab}C} \, \chi^D \,.
\end{equation}
Since in the latter equation the pair of coordinate indices $ab$ is skew-symmetric, it follows, by means of table \ref{Table spinors equivalent}, that this pair of vectorial indices can be replaced by one spinorial index up and one spinorial index down with vanishing contraction,
$$  (\nabla_a\,\nabla_b \me \nabla_b\,\nabla_a)\, \chi^C   \,\sim\, \mathfrak{R}^{A\ph{B}C}_{\ph{A}B\ph{C}D}\, \chi^D \quad \textrm{ where } \quad \mathfrak{R}^{A\ph{A}C}_{\ph{A}A\ph{C}D}\eq 0 \,.$$
Indeed, it is possible to check that equation (\ref{SpinorCurvature1}) is equivalent to the following relation
\begin{equation}\label{SpinorCurvature2}
  \varepsilon^{GABC}\,(\nabla_{AB}\,\nabla_{CD}\me \nabla_{CD}\,\nabla_{AB})\,\chi^E  \eq 4\,\mathfrak{R}^{G\ph{D}E}_{\ph{A}D\ph{E}F}\,\chi^F  \,.
\end{equation}
Inasmuch as the Levi-Civita connection is torsionless, the action of the curvature operator $2\nabla_{[a}\,\nabla_{b]}$ on the scalar $\chi^E\lambda_{E}$ must vanish. Then, once we have chosen the spinorial connection to obey the Leibniz rule with respect to the contraction of spinorial indices, one concludes that
\begin{equation}\label{SpinorCurvature3}
  \varepsilon^{GABC}\,(\nabla_{AB}\,\nabla_{CD}\me \nabla_{CD}\,\nabla_{AB})\,\lambda_E  \eq-\, 4\,\mathfrak{R}^{G\ph{D}F}_{\ph{A}D\ph{F}E}\,\lambda_F \,.
\end{equation}
It turns out that the curvature of the spinorial connection, in the gauge assumed here, can be entirely written in terms of the Riemann tensor associated to the metric $g_{ab}$. Indeed, for such a gauge we have that the spinorial curvature operator is given by $\frac{1}{4}R_{ab}^{\ph{ab}cd}\gamma_c \gamma_d$ \cite{Bat-PureSubspaces}, where $R_{ab}^{\ph{ab}cd}$ stands for the Riemann tensor and $\gamma_a$ are the six-dimensional Dirac matrices. Therefore, just as the Riemann tensor can be written as the sum of the of the Weyl tensor, the traceless part of the Ricci tensor and the Ricci scalar, the same can be done with $\mathfrak{R}^{G\ph{D}F}_{\ph{A}D\ph{F}E}$. According to table \ref{Table spinors equivalent}, the spinorial representation of the Weyl tensor is given by an object of the form $\Psi^{AB}_{\phantom{AB}CD}$  that is symmetric in both pairs of indices and traceless, while the  spinorial version of the traceless part of the Ricci tensor is given by an object of the form $\Phi^{AB}_{\phantom{AB}CD}$ that is skew-symmetric in both pairs of indices and traceless. Thus, if $R$ is the Ricci scalar, one can decompose that the spinorial curvature as follows
$$ 4\,\mathfrak{R}^{A\ph{C}B}_{\ph{A}C\ph{B}D} \eq \Psi^{AB}_{\phantom{AB}CD} \ma \Phi^{AB}_{\phantom{AB}CD} \ma R\left(\alpha\,\delta^A_C\,\delta^B_D  \me \beta\, \delta^A_D\,\delta^B_C\right) \,,$$
where
$$\Psi^{AB}_{\phantom{AB}CD}=\Psi^{(AB)}_{\phantom{AB}(CD)}\quad,\quad \Phi^{AB}_{\phantom{AB}CD}=\Phi^{[AB]}_{\phantom{AB}[CD]} \quad ,\quad \Psi^{AB}_{\phantom{AB}CB} \eq 0 \eq  \Phi^{AB}_{\phantom{AB}CB} \quad,$$
with $\alpha$ and $\beta$ being some constants. A relation between these constants can be determined imposing the requirement that $\mathfrak{R}^{A\ph{A}C}_{\ph{A}A\ph{C}D}$ vanishes, which yields $\beta=4\alpha$. Then, defining $\Lambda = \alpha \,R$, it follows that the spinorial curvature is given by
\begin{equation}\label{SpinCurv-Riemann}
   4\,\mathfrak{R}^{A\ph{C}B}_{\ph{A}C\ph{B}D} \eq \Psi^{AB}_{\phantom{AB}CD} \ma \Phi^{AB}_{\phantom{AB}CD} \ma \Lambda\left(\delta^A_C\,\delta^B_D  \me 4\, \delta^A_D\,\delta^B_C\right) \,,
\end{equation}
The above decomposition of the spinorial curvature have been used in \cite{Kerr6D} to find an integrability condition for the twistor equations in six-dimensional Einstein spaces.

Now, let  $V^{AB}=V^{[AB]}$ be a vector field. Then,
\begin{align*}
 2\, V^{AD}\,V_{BD} & \eq 2\,\lef \frac{1}{2}\, \varepsilon^{ADCE}\,V_{CE} \rig  \lef \frac{1}{2}\, \varepsilon_{BDGH}\,V^{GH} \rig
    \\
  & \eq 3\, \delta^{[A}_{B}\, \delta^{C}_{G}\, \delta^{E]}_{H}\, V_{CE} \,V^{GH} \eq  \delta^{A}_{B}\, V_{CD} \,V^{CD} \me 2\,  V^{AD}\,V_{BD}
\end{align*}
Therefore, if $V^{AB}$ is an arbitrary vector field then the following algebraic identity holds
\begin{equation}\label{VV}
  V^{AD}\,V_{BD} \eq \frac{1}{4} \, \delta^{A}_{B}\, \,V^{CD}\,V_{CD}  \,.
\end{equation}
Since the connection adopted here is torsionfree, it follows that the action of two covariant derivatives on an arbitrary scalar function $f$ commutes, so that Eq. (\ref{VV}) admits the following natural generalization
\begin{equation}\label{DD}
  \nabla^{AD}\, \nabla_{BD} \,f \eq \frac{1}{4} \, \delta^{A}_{B}\, \, \nabla^{CD}\, \nabla_{CD} \, f \eq  \frac{1}{4} \, \delta^{A}_{B}\, \,\square \, f \,.
\end{equation}
Where the box operator is defined by $\square = \nabla^{AB}\nabla_{AB}$. However, the latter identity does not hold when the derivative operators are acting on tensors or spinors, due to the fact that in these cases the covariant derivatives cease to commute. In particular, when the derivative operators act in spinors, one can prove that the correct analogs of Eq. (\ref{DD}) are given by the following relations
\begin{align}
  \nabla^{AC}\,\nabla_{BC} \chi^D \eq \frac{1}{4}\, \left( \delta^A_B \,\square\,\chi^D \me 4\, \mathfrak{R}^{A\ph{B}D}_{\ph{A}B\ph{B}E}\,\chi^E  \right)
  \quad ,& \quad \nabla_{BC}\,\nabla^{AC} \chi^D \eq \frac{1}{4}\, \left( \delta^A_B \,\square\,\chi^D \ma 4\, \mathfrak{R}^{A\ph{B}D}_{\ph{A}B\ph{B}E}\,\chi^E  \right)\,,\nonumber\\
  \label{Identity6D} \\
  \nabla^{AC}\,\nabla_{BC} \lambda_D \eq \frac{1}{4}\, \left( \delta^A_B \,\square\,\lambda_D \ma 4\, \mathfrak{R}^{A\ph{B}E}_{\ph{A}B\ph{B}D}\,\lambda_E  \right)
  \quad ,& \quad \nabla_{BC}\,\nabla^{AC} \lambda_D \eq \frac{1}{4}\, \left( \delta^A_B \,\square\,\lambda_D \me
   4\, \mathfrak{R}^{A\ph{B}E}_{\ph{A}B\ph{B}D}\,\lambda_E  \right) \,.\nonumber
\end{align}
These identities can be proved following steps entirely analogous to the ones used to obtain Eq. (\ref{VV}) and using Eqs. (\ref{SpinorCurvature2}) and (\ref{SpinorCurvature3}) to interchange the order of the covariant derivatives acting on the spinors. Likewise, these relations can be naturally generalized to the case in which the derivatives act in spinorial objects of higher rank. For instance,
$$ \nabla^{AC}\,\nabla_{BC}\,B^{D}_{\;\;F} \eq \frac{1}{4}\, \left( \delta^A_B \,\square\,B^{D}_{\;\;F} \me 4\, \mathfrak{R}^{A\ph{B}D}_{\ph{A}B\ph{B}E}\,B^{E}_{\;\;F} \ma  4\, \mathfrak{R}^{A\ph{B}E}_{\ph{A}B\ph{B}F} \,B^{D}_{\;\;E}  \right)\,. $$


\section{Conformal Transformations}\label{Sec.ConfTransf}

The goal of the present section is to obtain how the spinorial connection and its curvature are modified by a conformal transformation. If $g_{ab}$ denotes the metric, then a general conformal transformation amounts to the following map:
\begin{equation}\label{Conf_g}
 g_{ab}\;\mapsto\; \widetilde{g}_{ab}\,=\, \Omega^2\, g_{ab}\,,
\end{equation}
where $\Omega$ is a positive function throughout the manifold. Since in the spinorial formalism the metric is given by the totally skew-symmetric symbol $\varepsilon_{ABCD}$, one concludes that the following transformations hold
\begin{equation}\label{Conf_epsi}
  \widetilde{\varepsilon}_{ABCD}\,=\,\Omega^2 \,\varepsilon_{ABCD} \quad \textrm{ and }  \quad  \widetilde{\varepsilon}^{ABCD}\,=\,\Omega^{-2} \,\varepsilon^{ABCD} \,.
\end{equation}
Now, it is useful to define the following vector field
\begin{equation}\label{Def.n_H}
H_{AB}\,=\, \frac{1}{\Omega}\,\nabla_{AB}\Omega  \,.
\end{equation}
It is well-known that the Levi-Civita connections $\nabla_a$ and $\widetilde{\nabla}_{a}$ associated to $g_{ab}$ and $\widetilde{g}_{ab}$ respectively can be related by the following relation
$$ \widetilde{\nabla}_{a} V_b  \eq  \nabla_{a} V_b \me V_a\,H_b \me  V_b\,H_a  \ma g_{ab}\,V_c\,H^c  \,,   $$
whose spinorial analogue is given by
\begin{equation}\label{Added1}
  \widetilde{\nabla}_{AB} V_{CD}  \eq  \nabla_{AB} V_{CD} \me V_{AB}\,H_{CD} \me  V_{CD}\,H_{AB}  \ma \frac{1}{2}\,\varepsilon_{ABCD} V_{EF}\,H^{EF} \,.
\end{equation}
Then, writing $V_{AB} \eq \lambda_{[A}\,\xi_{B]}$ and assuming the relations
$$ \widetilde{\nabla}_{AB}\lambda_C = \nabla_{AB}\lambda_C  \ma 2 \, H_{C[A}\lambda_{B]}  \quad  \textrm{and} \quad
 \widetilde{\nabla}_{AB}\xi_C = \nabla_{AB}\xi_C  \ma 2 \, H_{C[A}\xi_{B]} \,,$$
 it is straightforward to prove that the identity (\ref{Added1}) is satisfied. Once we know how the connection $\widetilde{\nabla}_{AB}$ acts in a spinor of negative chirality, we can obtain its action on a spinor of positive chirality by using the fact that the connections  $\widetilde{\nabla}_{AB}$ and $\nabla_{AB}$ must agree when acting on a scalar. Then, choosing the scalar to be $\lambda_A\psi^A$, we are led to the identity
$$   \widetilde{\nabla}_{AB}\psi^ C  = \nabla_{AB}\psi^ C  \me 2 \,\psi^ D H_{D[A}\delta_{B]}^C \,. $$
Summing up, the action of the connection $\widetilde{\nabla}_{AB}$ on spinors is given by the following important relations:
\begin{align}
    \widetilde{\nabla}_{AB}\chi^ C  &= \nabla_{AB}\chi^ C  \me 2 \,\chi^ D H_{D[A}\delta_{B]}^C \,\,,\nonumber \\
    \quad \label{Conf_Connection1}\\
    \widetilde{\nabla}_{AB}\lambda_C &= \nabla_{AB}\lambda_C  \ma 2 \, H_{C[A}\lambda_{B]} \,\,. \nonumber
  \end{align}
Since both connections $\widetilde{\nabla}$ and $\nabla$  obey the Leibniz rule, one can use Eq. (\ref{Conf_Connection1}) to obtain the action of $\widetilde{\nabla}$ on objects of higher rank. For instance, it follows that
$$ \widetilde{\nabla}_{AB} \, Y^{CD}_{\ph{AB}EF} \eq  \nabla_{AB} \, Y^{CD}_{\ph{AB}EF} \me 2 \, Y^{GD}_{\ph{AB}EF}\, H_{G[A}\delta_{B]}^C
\me 2 \, Y^{CG}_{\ph{AB}EF}\, H_{G[A}\delta_{B]}^D \ma 2 \, H_{E[A}Y^{CD}_{\ph{AB}B]F} \ma 2 \, H_{F[A|}Y^{CD}_{\ph{AB}E|B]} \,,$$
where the vertical bars in the notation $H_{F[A|}Y^{CD}_{\ph{AB}E|B]}$ means that the index $E$ should not be anti-symmetrized along with $A$ and $B$. Since the property that defines the spinorial affine connection $\nabla_{AB}$ is that its action on $\varepsilon_{ABCD}$ yields zero, one can prove that the relations presented in Eq. (\ref{Conf_Connection1}) are indeed correct by computing $\widetilde{\nabla}_{AB} \widetilde{\varepsilon}_{CDEF}$ and noting that it vanishes.

Now, inserting (\ref{Conf_Connection1}) in the left hand side of the relation
$$  \widetilde{\varepsilon}^{\,GABC}\,(\widetilde{\nabla}_{AB}\,\widetilde{\nabla}_{CD}\me \widetilde{\nabla}_{CD}\,\widetilde{\nabla}_{AB})\,\chi^E  \eq 4\,\widetilde{\mathfrak{R}}^{G\ph{D}E}_{\ph{A}D\ph{E}F}\,\chi^F  \,,  $$
leads to the following transformation rules for the irreducible parts of the curvature:
\begin{align}
  \widetilde{\Psi}^{AB}_{\phantom{AB}CD} \eq&  \Omega^{-2}\,\Psi^{AB}_{\phantom{AB}CD} \,\,, \nonumber\\
  \widetilde{\Phi}^{AB}_{\phantom{AB}CD} \eq& \Omega^{-2}\,\left[ \Phi^{AB}_{\phantom{AB}CD} +
H^{AB}\,H_{CD} -  \nabla^{AB}H_{CD}  - \frac{1}{6} \,\,\delta^{[A}_C\delta^{B]}_D\, ( H^{EF}H_{EF}  - \nabla^{EF}H_{EF})   \right] \,\,, \label{CurvatureTransf}\\
\widetilde{\Lambda} \eq&  \Omega^{-2}\,\left(  \Lambda \ma \frac{1}{24}\,\nabla^{EF}H_{EF} \ma \frac{1}{12}\,H^{EF}H_{EF}  \right) \,\,. \nonumber
\end{align}
In particular, note that the object that represents the Weyl tensor, $\Psi^{AB}_{\phantom{AB}CD}$, transforms covariantly under conformal transformations, which stems from the fact that the Weyl tensor is the conformally invariant part of the Riemann tensor.


\section{Massless Fields}\label{Sec.MasslessFields}

In four dimensions, the prototype of a conformally invariant massless field equation is given by the source-free Maxwell's equations. These equations can be nicely cast in the language of differential forms by saying that the field strength tensor, $\bl{F}$, is an harmonic 2-form, namely $d\bl{F}=0$ and $d\star \bl{F}=0$, where $\star \bl{F}$ stands for the Hodge dual of $\bl{F}$. Given the relevance of these equations, the intent of the present section is to analyse how the harmonic\footnote{A differential form $\bl{F}$ is defined to be harmonic when $\Delta \bl{F} = ( d\delta + \delta d) \bl{F} = 0$, where
$\delta \propto \star d \star$ is the adjoint of the exterior derivative with respect to the usual inner product in the bundle of differential forms. Thus, every differential form that is closed and co-closed, namely $d\bl{F}=0$ and $d\star \bl{F}=0$, is harmonic, but the converse is not true in general. Nevertheless, in the particular case of a compact manifold endowed with a metric of Euclidean signature, these two concepts are equivalent. Here, we shall abuse the language and use the term harmonic to mean a closed and co-closed form, in any signature.} condition of a differential form is expressed in the six-dimensional spinorial formalism and investigate its conformal properties.

Let $\bl{B}$ be a 2-form. Then, the equation $d\bl{B}=0$ is equivalent to $\nabla_{[a}B_{bc]}=0$, which says that a 3-form made out of the derivative of $\bl{B}$ must vanish. In six dimensions, a 3-form  is represented in the spinorial language by the pair $(T_{AB}, T^{AB})$, where both objects are symmetric, $T_{AB}=T_{(AB)}$ and $T^{AB}=T^{(AB)}$\footnote{It is worth remembering that in general $T_{AB}$ is completely independent from $T^{AB}$, each one carry 10 degrees of freedom, summing the 20 degrees of a 3-form in six dimensions.}. Apart from a multiplicative constant, the only pair of this form that we can built using the derivative of $B$ in the spinorial language is $(\nabla_{C(A}B^C_{\phantom{C}B)}, \nabla^{C(A}B^{B)}_{\phantom{B)}C})$. From which we conclude that $d\bl{B}=0$ is equivalent to $\nabla_{C(A}B^C_{\phantom{C}B)}=0$ and $\nabla^{C(A}B^{B)}_{\phantom{B)}C}=0$. On the other hand, equation $d\star \bl{B}=0$ is equivalent to $\nabla^aB_{ab}=0$, \textit{i.e.} a 1-form made out of the derivative of $\bl{B}$ vanishes. In the spinorial formalism, a 1-form is represented by an object with a pair of skew-symmetric spinorial indices, $\omega_{AB}=\omega_{[AB]}$. One can also use the metric to raise this pair of indices, yielding $\omega^{AB}=\omega^{[AB]}$. There are two 1-forms that we can construct using the derivative of $B$ in the spinorial language, namely $\nabla_{C[A}B^C_{\phantom{C}B]}$ and $\varepsilon_{ABDE} \nabla^{CD}B^{E}_{\phantom{B}C}$. Using the fact that $B^A_{\phantom{A}A}=0$ it can be proved that these two 1-forms are, actually, proportional to each other, leading us to the conclusion that the spinorial equivalent of $d\star \bl{B}=0$ is $\nabla_{C[A}B^C_{\phantom{C}B]}=0$. Using the same reasoning we can arrive at the following relations:
\begin{align}
  \nonumber d \bl{\omega}\,&=\, 0 \quad \Leftrightarrow \quad \nabla_{CA}\omega^{CB}\,-\, \frac{1}{4}\delta^B_{\;A}\nabla_{CD}\omega^{CD}\,=\,0 \,,\\
  \nonumber d\star \bl{\omega} \,&=\, 0 \quad \Leftrightarrow \quad \nabla_{AB}\omega^{AB}\,=\,0 \,,\\
  \nonumber\\[-8pt]
  \nonumber d\bl{B}\,&=\, 0 \quad \Leftrightarrow \quad \nabla_{C(A}B^C_{\phantom{C}B)}\,=\,0\;\;\textrm{and}\;\; \nabla^{C(A}B^{B)}_{\phantom{B)}C}\,=\,0 \,,\\
   d \star \bl{B} \,&=\, 0 \quad \Leftrightarrow \quad \nabla_{C[A}B^C_{\phantom{C}B]}\,=\,0 \,, \label{HarmonicFields}\\
  \nonumber\\[-8pt]
  \nonumber d\bl{T}\,&=\, 0 \quad \Leftrightarrow \quad \nabla_{CA}T^{BC}\,-\,\nabla^{CB}T_{AC} \,=\,0 \,,\\
  \nonumber d\star \bl{T} \,&=\, 0 \quad \Leftrightarrow \quad \nabla_{CA}T^{BC}\,+\,\nabla^{CB}T_{AC} \,=\,0 \,.
\end{align}
Where in the above equations $\bl{\omega}$ is a 1-form, $\bl{B}$ is a 2-form and $\bl{T}$ is a 3-form.

Now, we shall investigate how these equations behave under conformal transformations and whether these equations are conformally invariant. Since we do not know what should be the conformal images of these three fields, let us define
$$\widetilde{\omega}^{AB}=\Omega^\sigma\,\omega^{AB} \quad,\quad \widetilde{B}^{A}_{\phantom{A}B} = \Omega^\kappa\, B^{A}_{\phantom{A}B} \quad,\quad \widetilde{T}_{AB} = \Omega^\tau \,T_{AB} \quad,\quad  \widetilde{T}^{AB} = \Omega^\varrho\, T^{AB}\,, $$
 where $\sigma$, $\kappa$, $\tau$ and $\varrho$ are free parameters that can be conveniently chosen. Using Eq. (\ref{Conf_Connection1}) along with the Leibniz rule, it is straightforward to see that for a general 1-form $\omega^{AB}=\omega^{[AB]}$, for a general 2-form $B^{A}_{\phantom{A}B}$, with $B^{A}_{\phantom{A}A}=0$, and for a general 3-form $(T^{AB}, T_{AB})= (T^{(AB)}, T_{(AB)})$,  the following identities are valid:
  \begin{align}
   \widetilde{\nabla}_{AB}\widetilde{\omega}^{AD} \eq & \Omega^{\sigma}\,\nabla_{AB}\omega^{AD} \ma(2+\sigma)\,H_{AB}\,\widetilde{\omega}^{AD}\ma \delta^D_{\;B} \,H_{AC}\,\widetilde{\omega}^{AC} \,,\nonumber \\
   \widetilde{\nabla}_{AB}\widetilde{B}^{A}_{\phantom{A}D} \eq & \Omega^{\kappa}\,\nabla_{AB}B^{A}_{\phantom{A}D} \ma (3+\kappa)\; H_{AB}\,\widetilde{B}^{A}_{\phantom{A}D}\me H_{AD}\,\widetilde{B}^{A}_{\phantom{A}B} \,,\nonumber\\
\Omega^2\,\widetilde{\nabla}^{AB}\widetilde{B}^{C}_{\phantom{C}A} \eq & \Omega^{\kappa}\,\nabla^{AB} B^{C}_{\phantom{C}A} \ma (3+\kappa)\; H^{AB}\,\widetilde{B}^{C}_{\phantom{C}A} \me H^{AC}\,\widetilde{B}^{B}_{\phantom{B}A}\,,\label{Derivatives1}\\
   \widetilde{\nabla}_{AB}\widetilde{T}^{CA} \eq &\Omega^{\rho}\,\nabla_{AB}\,T^{CA}  \ma  (4+\varrho)\;H_{AB}\,\widetilde{T}^{CA}\,,\nonumber\\
   \Omega^2\,\widetilde{\nabla}^{AB}\widetilde{T}_{CA} \eq & \Omega^{\tau}\,\nabla^{AB}\,T_{CA}  \ma (2+\tau)\;H^{AB}\,\widetilde{T}_{CA}\,. \nonumber
 \end{align}
Then, assuming the forms $\bl{\omega}$, $\bl{B}$ and $\bl{T}$ to be harmonic according to the metric $g_{ab}$, \textit{i.e.} assuming that (\ref{HarmonicFields}) holds, and using Eq. (\ref{Derivatives1}) we easily find that
\begin{align}
& \widetilde{\nabla}_{AB}\widetilde{\omega}^{AD}\me \frac{1}{4}\,\delta^D_{\;B}\widetilde{\nabla}_{AC}\widetilde{\omega}^{AC}  \eq (2+\sigma)\left[\,H_{AB}\, \widetilde{\omega}^{AD} \me \delta^D_{\;B} \;H_{AC}\,\widetilde{\omega}^{AC} \,\right] \,,\nonumber \\
 &\widetilde{\nabla}_{AB}\widetilde{\omega}^{AB} \eq  (6+\sigma)\,H_{AB} \,\widetilde{\omega}^{AB} \,, \nonumber  \\
  &\widetilde{\nabla}_{A(B}\widetilde{B}^{A}_{\phantom{A}D)}  \eq (2+\kappa)\,H_{A(B}\,\widetilde{B}^{A}_{\phantom{A}D)} \quad , \quad \widetilde{\nabla}^{A(B}\widetilde{B}^{C)}_{\phantom{C}A}  \eq  (2+\kappa)\,\Omega^{-2}\,H^{A(B}\,\widetilde{B}^{C)}_{\phantom{C}A} \,, \label{Derivatives2}\\
&\widetilde{\nabla}_{A[B}\widetilde{B}^{A}_{\phantom{A}D]} \eq (4+\kappa)\,H_{A[B}\,\widetilde{B}^{A}_{\phantom{A}D]} \nonumber\,, \\
&\widetilde{\nabla}_{CA}\widetilde{T}^{BC}\,-\,\widetilde{\nabla}^{CB}\widetilde{T}_{AC} \eq  (4+\varrho)\;H_{CA}\,\widetilde{T}^{BC} \me (2+\tau)\;\Omega^{-2}\,H^{CB}\,\widetilde{T}_{AC} \,, \nonumber \\
&\widetilde{\nabla}_{CA}\widetilde{T}^{BC}\,+\,\widetilde{\nabla}^{CB}\widetilde{T}_{AC} \eq  (4+\varrho)\;H_{CA}\,\widetilde{T}^{BC} \ma (2+\tau)\;\Omega^{-2}\,H^{CB}\,\widetilde{T}_{AC} \,.\nonumber
\end{align}
Comparing these equations with (\ref{HarmonicFields}), we conclude that the harmonic condition for the 1-form and for the 2-form are not conformally invariant. For instance, choosing $\kappa=-2$ we find that $d\widetilde{\bl{B}}=0$ while choosing $\kappa=-4$ we find that $d\tilde{\star}\widetilde{\bl{B}}=0$, but we cannot choose $\kappa$ in such a way that the both equations, $d\widetilde{\bl{B}}=0$ and $d\tilde{\star}\widetilde{\bl{B}}=0$, hold simultaneously for an arbitrary conformal transformation. In the same fashion, we cannot choose $\sigma$ in such a way that equations $d\bl{\omega}=0$ and  $d\star\bl{\omega}=0$ are simultaneously invariant under conformal transformations.  Differently, by choosing $\varrho=-4$ and $\tau=-2$ we see that both equations  $d\widetilde{\bl{T}}=0$ and $d\tilde{\star}\widetilde{\bl{T}}=0$ are conformally invariant. Thus, in six dimensions the harmonic condition is conformally invariant just for a 3-form. Actually, this fact is well-known and can be easily proved without using spinors, just using the fact that the exterior derivative is conformally invariant, since it does not depend on the metric, while the Hodge dual operator transforms homogeneously under conformal transformations. The reason for writing the above conditions in terms of spinors is to make connection with the upcoming section. Generally, in $n$ dimensions, the harmonic condition is conformally invariant just for $\frac{n}{2}$-forms.

Before proceeding, let us digress about the physical connection between massless fields and conformally invariant equations. The mass of a field provides a physical scale associated to the field, while massless fields have no intrinsic scale. Thus, since a conformal transformation provides a local change of scale,  we foresee a link between conformal invariance and massless fields. In addition, we know that massless particles move along null rays, \textit{i.e.} their motion are confined to the light cones. Then, since the light cone structure is conformally invariant, it is reasonable to expect that massless fields are described by conformally invariant equations. However, generally, this is not the case. For instance, if $\bl{L}$ is an harmonic $p$-form then the condition $d\bl{L}=0$ can be locally solved by writing $\bl{L}=d\bl{A}$, where the potential $\bl{A}$ is some $(p-1)$-form. In which case we have a gauge freedom of adding an exact form to the potential, $\bl{A}\rightarrow \bl{A} +d\bl{\Lambda}$, which leaves $\bl{L}$ unchanged. It turns out that one can manage to use this gauge freedom in order to make the equation $d\star\bl{L}=0$ equal to $\square \bl{A}\equiv \nabla^{a}\nabla_{a}\bl{A}=0$, which is clearly a massless equation. In this sense, the harmonic condition is a massless equation. In spite of this, as acknowledged above, the harmonic condition is conformally invariant just for middle forms, \textit{i.e.} just for $\frac{n}{2}$-forms, with $n$ being the dimension of the space. Another important example of a massless equation that is not conformally invariant is the equation for the gravitational field in general relativity. In a region of spacetime without matter we are left just with the gravitational field. In such a region, the  Ricci tensor vanishes, so that the Riemann tensor is equal to the Weyl tensor. Therefore, the Weyl tensor obeys the second Bianchi identity, $\nabla_{[a}C_{bc]de}=0$. However, if we perform a general conformal transformation in the metric, the Ricci tensor becomes non-vanishing and the Weyl tensor ceases to obey the second Bianchi identity. The examples of this paragraph illustrate that conformally invariant equations are much more special than massless field equations. Although it is simple matter to find massless field equations, conformally invariant equations are rare and hard to find. Indeed, there is a whole formalism, the tractor calculus, whose main application is the search and the study of conformally invariant differential operators \cite{Tractor}.

However, it is worth noting that it is always possible to view a massless field that is not conformally invariant as the byproduct of a larger conformally invariant massless theory with broken conformal symmetry \cite{Gover-OriginMass}. For instance, by considering the conformal factor $\Omega$ as a dynamical field in the gravitational theory, along with the metric, one arrives at a larger theory whose field equation is conformally invariant. For example, considering a theory whose fields are $(g_{ab}, \Omega)$ and whose field equation is $\mathrm{Ricci}[\Omega^2 g_{ab}]  = 0$,  it is obvious that such a theory is invariant under the conformal transformation  $(g_{ab}, \Omega)\rightarrow (\Theta^2\,g_{ab}, \Theta^{-1}\,\Omega)$. The latter theory have been introduced, with advantages, in Ref. \cite{Friedrich} with aim of studying the initial value problem in General Relativity.

As an aside, it is pertinent commenting on the relation between the harmonic condition and the source-free Yang-Mills equation. In gauge theory, if $\bl{A}$ is a connection of a principal bundle, namely a Lie algebra-valued 1-form, then its curvature is a Lie algebra-valued 2-form obeying Yang-Mills equations. In absence of sources, these equations are analogous to the harmonic condition, but with the exterior derivative replaced by a suitable covariant derivative $D$.  Then, in four dimensions, a prompt solution for Yang-Mills equations is provided by a 2-form $\bl{F}$ that is \emph{closed} and (anti-)self-dual, namely $D\bl{F}=0$ and $\star\bl{F}\propto \bl{F}$. This kind of solution is called an instanton and its non-perturbative nature makes it of great physical interest, specially in non-abelian gauge theory \cite{YangMills-Atiyah}. Nevertheless, in higher dimensions, self-dual 2-forms do not exist and instantons solutions for the Yang-Mills equations are not possible in the usual sense. However, there is a sophisticated mathematical treatment that enable the generalization of the concept of an instanton to six dimensions, the gerbe formalism \cite{gerbe_Witten}. Roughly speaking, a gerbe is a generalization of the notion of a principal bundle such that its connection is a $2$-form rather than a 1-form \cite{gerbe_Baez}. Thus, in six dimensions, we can say that a \emph{closed} (anti-)self-dual 3-form is the curvature of a connection on a gerbe that extremizes the action\footnote{Note that if $\bl{T}$ is a 3-form then $\bl{T}\wedge \bl{T}$ vanishes identically. Thus, if $\bl{T}$ is (anti-)self-dual then $\bl{T}\wedge \star \bl{T}$ vanishes. Moreover, in a real six-dimensional space of Euclidean signature the action $S=\int\bl{T}\wedge \star\bl{T}$ is positive definite. So, it would follow that a real instanton-like solution, $D\bl{T}=0$ and $\star \bl{T}\propto \bl{T}$, would be an absolute minimum of the action $S$ in Euclidean signature. However, it turns out that in a six-dimensional Euclidean space there is no real 3-form such that $\star\bl{T}\propto \bl{T}$, which stems from the fact that in such a case $\star (\star \bl{T})=-\bl{T}$. Differently, instanton-like solutions are allowed in the Lorentzian signature. }
$$ S \eq \int\,\textrm{Tr}( \,D\bl{A}\wedge \star D\bl{A} \,) \,,$$
where $\bl{A}$ is a Lie 2-algebra-valued 2-form \cite{gerbe_Mathai,gerbe_Witten}. Particularly, in string theory, the Kalb-Ramond 2-form can be seen as the connection of a $U(1)$ gerbe \cite{gerbe_Baez}. Specially in six dimensions, the fact that the harmonic condition for a 3-form is conformally invariant, while the same is not true for a 2-form, enhance the geometrical relevance of the gerbe formalism in this particular dimension. The geometric interpretation of a gerbe connection is quite natural: just as a 1-form connection couples to the vector field tangent to the orbit of a point particle to give the parallel transport of its internal degrees of freedom, a 2-form connection couples to the world-sheet of a string to provide its parallel transport.

\section{Conformally Invariant Spinorial Equations}\label{Sec.ConformallyIvariantEq}

The aim of this section is to present a series of conformally invariant equations that emerge in the six-dimensional spinorial formalism. The strategy to obtain these conformally invariant equations is to write down, using the index notation, three well-known conformally invariant equations, the massless Dirac equation, the twistor equation, and the equation that defines the integrability of a maximally isotropic distribution, and try to generalize these equations to higher order spinorial objects.


\subsection{Generalizing the Dirac Equation}

The Dirac operator is formally defined by $\mathcal{D}= \bl{e}_{\hat{a}}\cdot\,\nabla^{\hat{a}}$, where the dot represents the Clifford action of a vector field on a spinor, and $\{\bl{e}_{\hat{a}}\}$ is a local frame for the tangent bundle. In terms of the index notation in six dimensions, the Clifford action of a vector field on a Dirac spinor $\psi=(\psi^A,\psi_B)$ is given by
\begin{equation}\label{CliffordAction}
  \bl{V}\cdot \psi \eq ( \,2\,V^{AB}\,\psi_B\,,\, -\,2\,V_{BA}\,\psi^A \, ) \,.
\end{equation}
Indeed, one can check that this action leads to the relation that defines the Clifford algebra, namely
$$ \bl{e}_{\hat{a}}\cdot\left( \bl{e}_{\hat{b}}\cdot \psi \right) \ma \bl{e}_{\hat{b}}\cdot\left(\bl{e}_{\hat{a}}\cdot \psi \right) \eq 2\,g_{\hat{a}\hat{b}}\,\,\psi \,\,.$$
Then, by means of (\ref{CliffordAction}), one easily obtain the following equivalence relation
\begin{equation}\label{DiracEq}
 \mathcal{D}\,\psi \eq 0    \quad \Leftrightarrow \quad    \nabla^{AB}\,\psi_B \eq 0 \quad \textrm{and} \quad \nabla_{AB}\,\psi^B \eq 0 \,.
\end{equation}
The index form of the massless Dirac equation in six dimensions resembles its 4-dimensional correspondent, which can be cast in the form $\nabla^{\alpha\dot{\beta}}\xi_{\dot{\beta}}=0$ and  $\nabla_{\alpha\dot{\beta}}\zeta^{\alpha}=0$, where $\alpha$ and $\dot{\beta}$ are 4-dimensional spinorial indices ranging from 1 to 2, and the pair $(\zeta^{\alpha}, \xi_{\dot{\beta}})$ is a Dirac spinor in four dimensions.

Since the massless Dirac equation, $\mathcal{D}\psi = 0$, is conformally invariant, we conclude that both equations on the right hand side of (\ref{DiracEq}) might also be conformally invariant. Indeed, choosing the transformation of the Dirac spinor to be
$$ \psi\eq (\,\psi^A\,,\,\psi_B\,)  \quad \longrightarrow \quad   \widetilde{\psi}\eq (\,\widetilde{\psi}^A\,,\,\widetilde{\psi}_B\,) \eq (\,\Omega^{-3}\,\psi^A\,,\,\Omega^{-2}\,\psi_B\,) \, $$
and using (\ref{Conf_Connection1}), we can check that whenever the equations $\nabla^{AB}\psi_B = 0$ and  $\nabla_{AB}\psi^B = 0$ hold, the equations $\widetilde{\nabla}^{AB}\widetilde{\psi}_B = 0$ and  $\widetilde{\nabla}_{AB}\widetilde{\psi}^B = 0$ also hold, for any choice of conformal factor.

Now, summing and subtracting the last pair of equations in (\ref{Derivatives2}), it follows that the spinorial equations
\begin{equation}\label{ConfInv3vec}
 \nabla_{AB}\,T^{BC}\eq 0  \quad\textrm{ and }\quad \nabla^{AB}\,T_{BC}\eq 0 \,,
\end{equation}
for $T^{AB}=T^{(AB)}$ and $T_{AB}=T_{(AB)}$, are also conformally invariant if we choose the conformal transformation of the objects $T^{AB}$ and $T_{AB}$ to be
$$ T^{AB} \quad \longrightarrow \quad \widetilde{T}^{AB}\eq \Omega^{-4}\,T^{AB} \quad \textrm{ and } \quad  T_{AB} \quad \longrightarrow  \quad \widetilde{T}_{AB} \eq \Omega^{-2}\,T_{AB} \,.$$
Then, inspired by the conformally invariant equations (\ref{DiracEq}) and (\ref{ConfInv3vec}), it is natural to guess that the general equations
\begin{equation}\label{ConfInv1}
  \nabla_{AB_1}\,K^{B_1\ldots B_p} \eq 0 \quad \textrm{ and } \quad   \nabla^{AB_1}\,K_{B_1\ldots B_p} \eq 0
\end{equation}
are conformally invariant if these spinorial objects are totally symmetric, $K^{B_1\ldots B_p}= K^{(B_1\ldots B_p)}$ and $K_{B_1\ldots B_p}=K_{(B_1\ldots B_p)}$. Indeed, choosing the transformation rules
$$ K^{B_1\ldots B_p} \quad \longrightarrow \quad \widetilde{K}^{B_1\ldots B_p}\eq \Omega^{-(2+p)}\,K^{B_1\ldots B_p} \quad \textrm{ and } \quad
K_{B_1\ldots B_p} \quad \longrightarrow \quad \widetilde{K}_{B_1\ldots B_p}\eq \Omega^{-2}\,K_{B_1\ldots B_p} \,,$$
and using (\ref{Conf_Connection1}) along with its higher rank generalizations, it follows that
$$ \nabla_{AB_1}\,K^{B_1\ldots B_p} \eq 0 \quad \Leftrightarrow \quad \widetilde{\nabla}_{AB_1}\,\widetilde{K}^{B_1\ldots B_p} \eq 0\quad \textrm{ and } \quad   \nabla^{AB_1}\,K_{B_1\ldots B_p} \eq 0 \quad \Leftrightarrow \quad \widetilde{\nabla}^{AB_1}\,\widetilde{K}_{B_1\ldots B_p} \eq 0 \,,$$
for any choice of conformal factor $\Omega$. Actually, these equations have already been considered before in the context of the vector space $\mathbb{C}^6$ in Ref. \cite{Mason:2011nw}, but the conformal transformation of the fields $K^{B_1\ldots B_p}$ and $K_{B_1\ldots B_p}$ have not been pointed out there. The geometrical interpretation of these equations is, certainly, cumbersome  in the case of arbitrary $p$. However, the special case $p=2$ can be readily interpreted with the help of Eq. (\ref{HarmonicFields}). More precisely, the equation $\nabla_{AB}\,K^{BC}=0$ is equivalent to the assertion that the self-dual 3-form whose spinorial representation is given by $(K^{AB},0)$ is harmonic. Analogously, equation $\nabla^{AB}\,K_{BC}=0$ says that the anti-self-dual 3-form represented by $(0,K_{AB})$ is harmonic. Particularly, in the case in which $K^{AB}=\xi^A\xi^B$ for some nonzero spinor $\xi^A$, it follows that the 3-form $(K^{AB},0)$ generates a vector distribution, spanned by the vector fields annihilated by the interior product with this 3-form, that is maximally isotropic and whose associated pure spinor is $\xi^A$ \cite{Bat-Spin6D,GualtieriThesis,Bat-PureSubspaces}. In this case, equation $\nabla_{AB}\,K^{BC}=0$ is equivalent to the following condition
\begin{equation}\label{MaxIsotrpic}
  V^{AB}\,\nabla_{AB}\,\xi^C \,\propto \, \xi^C \quad \;\; \forall \quad V^{AB} \; \textrm{ of the form } \;  V^{AB}\eq \xi^{[A}\,\chi^{B]} \,.
\end{equation}
Since the vectors of the form $V^{AB}= \xi^{[A}\,\chi^{B]}$ are just the vectors annihilated by the 3-form $(\xi^A\xi^B,0)$, it follows that equation $\nabla_{AB}\,(\xi^B\xi^C)=0$  means that the isotropic distribution generated by the pure spinor $\xi^A$ is integrable.  Analogously,  equation $\nabla^{AB}\,(\kappa_B\kappa_C)=0$ is equivalent to the condition that the maximally isotropic distribution generated by the pure spinor $\kappa_A$ is integrable.\footnote{It is worth recalling that in even dimensions less than eight, every chiral spinor is a pure spinor.} The index form of the conformally invariant equations (\ref{ConfInv1}) shows a striking resemblance with the conformally invariant massless field equations studied by Penrose in four dimensions \cite{Peeling-Penrose}, which are given by $\nabla^{\alpha_1\dot{\beta}} \phi_{\alpha_1\cdots \alpha_p} = 0$ and  $\nabla^{\alpha\dot{\beta}_1}  \phi_{\dot{\beta}_1\cdots \dot{\beta}_p} = 0$, where $\phi_{\alpha_1\cdots \alpha_p}$ and $\phi_{\dot{\beta}_1\cdots \dot{\beta}_p}$ are totally symmetric in their spinorial indices.


\subsection{Generalizing the Twistor Equation}

Another well-known conformally invariant equation is the twistor equation. In $n$ dimensions, a Dirac spinor $\psi$ is called a twistor whenever it obeys the following differential constraint:
$$  \nabla_{\hat{a}}\,\psi \me \frac{1}{n}\, \bl{e}_{\hat{a}}\cdot\mathcal{D}\,\psi \eq 0 \,.   $$
Now, let us deduce how this equation is written in six dimensions using the index formalism. First, note that the twistor equation do not mix the two chiral parts of the Dirac spinor $\psi$. So, for sake of simplicity, let us suppose that the spinor $\psi$ is chiral. For instance, consider the case  of positive chirality, $\psi=(\psi^A,0)$. Then, due to (\ref{CliffordAction}), the twistor equation is written as
\begin{equation}\label{twistor2}
   e_{\hat{a}}^{\;BC}\,\nabla_{BC} \,\psi^A \me
\frac{1}{6}\,\left(  -\,4\, e_{\hat{a}}^{\;AF} \,e^{\hat{b}}_{\;FD} \,e_{\hat{b}}^{\;BC}  \right) \nabla_{BC}\, \psi^D  \eq 0 \,.
\end{equation}
Thus, contracting the latter equation with $e^{\hat{a}}_{\;GH}$ and using the identity
$$  e^{\hat{a}}_{\;AB} \,e_{\hat{a}}^{\;CD} \eq \delta_{A}^{\;[C} \delta_{B}^{\;D]}\,,$$
we end up with
\begin{equation}\label{twistor3}
   \nabla_{GH}\, \psi^A \eq \frac{1}{3} \, \left(\, \delta_{H}^{\;A}\,\nabla_{GD}\, \psi^D \me  \delta_{G}^{\;A}\,\nabla_{HD}\, \psi^D   \,\right)  \,.
\end{equation}
Then, in six dimensions, a positive chirality spinor $\psi^A$ is a twistor if, and only if, it obeys the equation
\begin{equation}\label{twistor+}
   \nabla_{AB}\,\psi^C \eq \zeta_{[A}\, \delta_{B]}^{\;C}
\end{equation}
for some spinor $\zeta_A$. Indeed, contracting a pair of indices in this equation one concludes that $\zeta_A$ is necessarily given by $\frac{2}{3} \nabla_{AB}\psi^B$, which agrees with (\ref{twistor+}). Differently from the definition adopted in \cite{Kerr6D}, note that the above deduction shows that the contraction $\psi^A\zeta_A$ is not necessarily zero. The special case $\psi^A\zeta_A=0$ represents the particular circumstance in which the twistor $\psi^A$ is an integrable pure spinor, namely the null subspaces generated by the vector fields that annihilate the pure spinor $\psi^A$ form an integrable foliation.

Analogously, a spinor of negative chirality, $\psi=(0,\psi_A)$, is a twistor if, and only if, there exits a spinor $\chi^A$ such that
\begin{equation}\label{twistor-}
   \nabla^{AB}\,\psi_C \eq \chi^{[A}\, \delta^{\;B]}_{C}  \,.
\end{equation}
Generally, if $\psi$ is an arbitrary Dirac spinor, $\psi=(\psi^A,\psi_B)$, then it will be a twistor if, and only if, both equations (\ref{twistor+}) and (\ref{twistor-}) are satisfied for some spinors $\zeta_A$ and $\chi^B$.   Raising and lowering the derivative indices in Eqs. (\ref{twistor+}) and (\ref{twistor-}) respectively, in accordance with (\ref{spinorialmetric}), we conclude that the twistor equation can be equivalently written as
\begin{equation}\label{Twistor-skew}
   \nabla^{AB}\,\psi^C \eq  \nabla^{[AB}\,\psi^{C]}   \quad\;\textrm{ and }\;\quad \nabla_{AB}\,\psi_C \eq  \nabla_{[AB}\,\psi_{C]}  \,.
\end{equation}
Moreover, in view of the skew-symmetry in the indices of the derivative operator, it is immediate to verify that these equations are also equivalent to the following pair of equations:
\begin{equation}\label{Twistor-symm}
   \nabla^{A(B}\,\psi^{C)} \eq  0   \quad\;\textrm{ and }\;\quad \nabla_{A(B}\,\psi_{C)} \eq  0  \,.
\end{equation}
Since the twistor equation is conformally invariant, it follows that equations (\ref{twistor+}), (\ref{twistor-}), (\ref{Twistor-skew}) and (\ref{Twistor-symm}) are all invariant under conformal transformations. In order to attain conformal invariance, the conformal transformation of the twistor components might be
$$ \psi^A \; \longrightarrow  \;  \widetilde{\psi}^A \eq \psi^A  \quad\;,\; \quad
 \psi_A \; \longrightarrow  \;  \widetilde{\psi}_A \eq \Omega\, \psi_A  \,.$$
It is interesting to compare the index form of the twistor equation in six and four dimensions. In the latter case, the twistor equations can be written as $\nabla_{\dot{\beta}(\alpha_1} \phi_{\alpha_2)} = 0$ and $\nabla_{\alpha (\dot{\beta}_1} \phi_{\dot{\beta}_2)} = 0$, which is quite similar to Eq. (\ref{Twistor-symm}). These striking resemblances between the spinorial formalism in four and six dimensions should not be expected in principle, inasmuch as in different dimensions the spinorial indices have different group-theoretical roots. Indeed, for example, Eq. (\ref{Twistor-skew}) have no analog in four dimensions, since the covariant derivative in such a case carry two spinorial indices of different types that cannot be treated in same fashion and, therefore, cannot be permuted. Moreover, in four dimensions,  the anti-symmetrization of three or more spinorial indices is identically zero, which would give the four-dimensional analog of  Eq. (\ref{Twistor-skew}) a trivial meaning.


The nice thing about the form (\ref{Twistor-symm}) of expressing the twistor equation is that it allows a higher rank generalization that is also conformally invariant. Indeed, assuming that $Q^{A_1\cdots A_p}$ and $Q_{A_1\cdots A_p}$ are totally symmetric, namely $Q^{A_1\cdots A_p}=Q^{(A_1\cdots A_p)}$ and $Q_{A_1\cdots A_p}=Q_{(A_1\cdots A_p)}$, it turns out that the equations
\begin{equation}\label{ConfInv2}
  \nabla^{A(B}\,Q^{C_1\cdots C_p)} \eq 0 \quad\;\textrm{ and }\;\quad   \nabla_{A(B}\,Q_{C_1\cdots C_p)} \eq 0
\end{equation}
are both conformally invariant if we choose the fields $Q^{A_1\cdots A_p}$ and $Q_{A_1\cdots A_p}$ to behave under conformal transformations as follows
$$  Q^{A_1\cdots A_p}\; \longrightarrow  \;  \widetilde{Q}^{A_1\cdots A_p} \eq Q^{A_1\cdots A_p}  \quad\;,\; \quad
 Q_{A_1\cdots A_p}\; \longrightarrow  \;  \widetilde{Q}_{A_1\cdots A_p} \eq \Omega^p\,Q_{A_1\cdots A_p}  \,, $$
where it is necessary to use (\ref{Conf_Connection1}) and its higher rank versions in order to verify this assertion. Just as Eq. (\ref{ConfInv1}) is a natural higher rank generalization of the Dirac equation, Eq. (\ref{ConfInv2}) is a higher rank generalization of the twistor equation. As far as the author knows, these equations have not been found elsewhere.  It is worth stressing that, generally, solutions of (\ref{ConfInv1}) do not generate solutions of (\ref{ConfInv2}) and \textit{vice versa}. This is in accordance with the fact that the Dirac and the twistor operators are complementary parts of the covariant derivative operator with respect to the Clifford action of a frame $\{\bl{e}_{\hat{a}}\}$, as depicted below.
$$  \bl{e}^{\hat{a}} \cdot \nabla_{\hat{b}} \; \longrightarrow  \;  \left\{ \begin{array}{l}
                    \textrm{Trace part:}\; \mathcal{D} \,\equiv\,  \bl{e}^{\hat{a}}\cdot \nabla_{\hat{a}}   \\
                   \\
                   \textrm{Traceless part:}\; \bl{e}^{\hat{a}}  \cdot \mathcal{T}_{\hat{b}} \,\equiv\, \bl{e}^{\hat{a}} \cdot \left( \nabla_{\hat{b}} \me \frac{1}{n}\, \bl{e}_{\hat{b}}\cdot \mathcal{D} \right)
                  \end{array}  \right.
  $$
Where by traceless it is meant that $\bl{e}^{\hat{a}} \cdot \mathcal{T}_{\hat{a}} = 0$. It is pertinent to highlight the power of the index notation, as it would be much more difficult to obtain the conformally invariant equations (\ref{ConfInv1}) and (\ref{ConfInv2}) without using the index formalism adopted here.

Now, we shall try to provide a geometrical interpretation for the objects $Q^{A_1\cdots A_p}$ and $Q_{A_1\cdots A_p}$  obeying (\ref{ConfInv2}). Let $\xi^A$ and $\kappa_A$ be integrable pure spinors that are parallel propagated along the curves tangent to their respective isotropic foliations, namely $\xi^A\nabla_{AB}\xi^C=0$ and $\kappa_A\nabla^{AB}\kappa_C=0$. Then, if $Q^{A_1\cdots A_p}$ and $Q_{A_1\cdots A_p}$ obey Eq. (\ref{ConfInv2}), it follows that the scalars $Q^{A_1\cdots A_p}\kappa_{A_1}\cdots \kappa_{A_p}$ and $Q_{A_1\cdots A_p}\xi^{A_1}\cdots \xi^{A_p}$ are constant along the foliations generated by $\kappa_A$ and $\xi^A$ respectively, namely
$$ \kappa_A\,\nabla^{AB}\, \left( Q^{A_1\cdots A_p}\kappa_{A_1}\cdots \kappa_{A_p} \right) \eq 0   \quad \; \textrm{and} \; \quad
  \xi^A\,\nabla_{AB}\, \left(  Q_{A_1\cdots A_p}\xi^{A_1}\cdots \xi^{A_p} \right)  \eq  0  \,.$$
In this sense, we can interpret the objects $Q^{A_1\cdots A_p}$ and $Q_{A_1\cdots A_p}$ as generators of conserved scalars along maximally isotropic foliations.



\subsection{Generalizing the Equation Defining an Integrable Maximally Isotropic Distribution}

Isotropic structures are, by definition, formed by null vector fields, which are conformally invariant. Therefore,  it is reasonable to expect that the equations that define their integrability are also conformally invariant. For instance, if $\xi^A$ is a nonzero spinor of positive chirality then the vectors fields that annihilate this spinor field form a maximally isotropic distribution, namely they span a null subspace of dimension three at each tangent space of the manifold. These vector subspaces form an integrable distribution in the sense of Frobenius  if, and only if,
$$  \xi^A\,\zeta^B\,\nabla_{AB}\,\xi^C \,\propto\, \xi^C \;,\quad \forall \quad \zeta^B \,.$$
Such condition is equivalent to the existence of some vector field $V_{AB}\eq V_{[AB]}$ such that
\begin{equation}\label{IntegrablePure+}
 \xi^A\,\nabla_{AB}\,\xi^C \eq  V_{AB}\, \xi^A \, \xi^C \,.
\end{equation}
In the same fashion, a spinor field of negative chirality, $\kappa_A$, is said to generate an integrable maximally isotropic distribution if, and only if, there exists  some vector field $\mathcal{V}^{AB}\eq \mathcal{V}^{[AB]}$ such that
\begin{equation}\label{IntegrablePure-}
\kappa_A\,\nabla^{AB}\,\kappa_C \eq   \mathcal{V}^{AB}\,\kappa_A \, \kappa_C \,.
\end{equation}
As expected, one can check that Eqs. (\ref{IntegrablePure+}) and (\ref{IntegrablePure-}) are conformally invariant,  with the spinors $\xi^A$ and $\kappa_A$ transforming as:
$$ \xi^A \;\longrightarrow \;  \widetilde{\xi}^A \eq \Omega^{-1}\, \xi^A  \quad\textrm{ and }\quad
 \kappa_A \;\longrightarrow \;  \widetilde{\kappa}_A \eq \, \kappa_A   \,.   $$
Now, natural higher rank generalizations of Eqs. (\ref{IntegrablePure+}) and (\ref{IntegrablePure-}) are provided by the following equations:
\begin{align}
L^{A_1(A_2\cdots A_p}\,\nabla_{A_1 B}\,L^{C_1\cdots C_p)}\eq&\,   V_{A_1 B}\, L^{A_1(A_2\cdots A_p}\,L^{C_1\cdots C_p)} \,, \nonumber\\
  L_{A_1(A_2\cdots A_p}\,\nabla^{A_1 B}\,L_{C_1\cdots C_p)}\eq&\,   \mathcal{V}^{A_1 B}\, L_{A_1(A_2\cdots A_p}\,L_{C_1\cdots C_p)}   \label{ConfInvL} \,.
\end{align}
Where $V_{AB}=V_{[AB]}$ and $\mathcal{V}^{AB}=\mathcal{V}^{[AB]}$ are some vector fields, whereas $L^{A_1\cdots A_p}= L^{(A_1\cdots A_p)}$ and $L_{A_1\cdots A_p}= L_{(A_1\cdots A_p)}$ are totally symmetric. Using (\ref{Conf_Connection1}) and its higher rank generalizations, one can check that the quadratic equations (\ref{ConfInvL}) are also conformally invariant if the latter fields transform as
$$ L^{A_1\cdots A_p} \;\longrightarrow \;  \widetilde{L}^{A_1\cdots A_p} \eq \Omega^{-p}\, L^{A_1\cdots A_p} \quad\textrm{ and }\quad
 L_{A_1\cdots A_p} \;\longrightarrow \;  \widetilde{L}_{A_1\cdots A_p} \eq \, L_{A_1\cdots A_p}  \,.   $$
To the best of author's knowledge, these equations have not been considered elsewhere. Now, let us try to give some sort of interpretation to these conformally invariant equations. Let $\psi^A$ and $\psi_{A}$ be twistors, namely Eq. (\ref{Twistor-skew}) hold. Then, defining $\eta^A=L^{A B_2\cdots B_p}\psi_{B_2}\cdots \psi_{B_p}$ and $\eta_A=L_{A B_2\cdots B_p}\psi^{B_2}\cdots \psi^{B_p}$ it follows that the scalars $\eta^A\psi_A$ and $\eta_A\psi^A$ are such that
$$ \eta^A\,\left(\nabla_{AB} \me  V_{AB} \right) \,\left( \eta^C\psi_C \right) \eq 0 \quad\textrm{ and }\quad   \eta_A\,\left(\nabla^{AB} \me  \mathcal{V}^{AB} \right) \, \left( \eta_C\psi^C \right) \eq 0  \,. $$
Particularly, in the case $p=1$ and $V_{AB}=0$ the first condition above states that $L^A\nabla_{AB}(L^C\psi_C)=0$, which means that the scalar $L^A\psi_A$ is constant along the foliation generated by the integrable pure spinor $L^A$, for any twistor $\psi_A$.

It is worth pointing out that the dimension six is the only even dimension in which there exist pure spinors that are twistors but do not generate an integrable distribution \cite{Bat-PureSubspaces}. For instance, in four dimensions, the twistor equation for a Weyl spinor of positive chirality is given by $\nabla_{\dot{\beta}(\alpha_1} \phi_{\alpha_2)} = 0$, which can be equivalently written as $\nabla_{\dot{\beta}\alpha_1} \phi_{\alpha_2} = \varepsilon_{\alpha_1 \alpha_2}\,\lambda_{\dot{\beta}}$, for some spinor $\lambda_{\dot{\beta}}$. The latter equation, in turn, implies that  $\phi^{\alpha_1}\nabla_{\dot{\beta}\alpha_1} \phi_{\alpha_2} = \phi_{\alpha_2}\,\lambda_{\dot{\beta}}$, which means that the isotropic distribution generated by $\phi_\alpha$ is integrable. Actually, in four dimensions, a Weyl spinor generates an integrable distribution of isotropic planes if, and only if, it obeys the twistor equation \cite{Bat-PureSubspaces}.


\section{Some Integrability Conditions}\label{Sec.IntegrabilityCond}

The goal of this section is to present some integrability conditions that must be satisfied by the curvature of the Levi-Civita connection in order for the conformally invariant equations exhibited in the preceding section to admit solutions. More precisely, we shall calculate the first order integrability conditions. Although for the Eqs. (\ref{ConfInv2}) and   (\ref{ConfInvL}) we are going to consider just the simplest case $p=1$, these constraints already give a hint in the kind of integrability conditions that are behind such differential equations.

\subsection{Dirac Equation and its Generalizations}

Let us start analysing the integrability conditions of the massless Dirac equation and its generalizations. If $K^{A_1\cdots A_p}$ and  $K_{A_1\cdots A_p}$ are objects that are totally symmetric in its indices, then the identity (\ref{Identity6D})  yields
\begin{align}
  4\,\nabla^{AD}\,\nabla_{BD} K^{C_1\cdots C_p} \eq &  \delta^A_B \,\square\,K^{C_1\cdots C_p} \me 4\,p\,\, \mathfrak{R}^{A\ph{B}(C_1}_{\ph{A}B\ph{(C}E}\,K^{C_2\cdots C_p)E}
  \nonumber\,,\\
   \label{Identity6D2}\\
  4\, \nabla_{AD}\,\nabla^{BD} K_{C_1\cdots C_p}\eq &  \delta^B_A \,\square\,K_{C_1\cdots C_p} \me
   4\,p\,\, \mathfrak{R}^{B\ph{B}E}_{\ph{B}A\ph{B}(C_1}\,K_{C_2\cdots C_p) E} \nonumber\,.
\end{align}
Now, suppose that $K^{A_1\cdots A_p}$ and  $K_{A_1\cdots A_p}$ obey the conformally invariant equations (\ref{ConfInv1}). Then, contracting the indices $B$ and $C_1$ in (\ref{Identity6D2}) yields
\begin{equation}\label{SquareK}
 \square\,K^{A C_2\cdots C_p} \eq 4\,p\,\, \mathfrak{R}^{A\ph{B}(B}_{\ph{A}B\ph{(C}E}\,K^{C_2\cdots C_p)E} \quad\textrm{ and }\quad
\square\,K_{A C_2\cdots C_p} \eq  4\,p\,\, \mathfrak{R}^{B\ph{B}E}_{\ph{B}A\ph{B}(B}\,K_{C_2\cdots C_p) E} \,.
\end{equation}
In the case $p=1$, \textit{i.e.} for the case of spinors obeying the massless Dirac equation, these relations give
$$ \square \, K_A \eq -\,15\,\Lambda\, K_A   \quad\textrm{ and }\quad   \square \, K^A \eq -\,15\,\Lambda\, K^A \,. $$
Nevertheless, such relations impose no constraint on the curvature. Moreover, analogous relations are valid in any dimension, as a consequence of the following well-known identity \cite{Bat-PureSubspaces}:
$$ \mathcal{D}  \mathcal{D} \,\psi \eq \square\,\psi  \ma \frac{1}{4}\,R \,\psi \,,$$
where $\psi$ is an arbitrary spinor and $R$ is the Ricci scalar. Where it is pertinent to recall that in the present formalism the scalar $\Lambda$ is equal to $R$ apart from a multiplicative constant. In the case $p=2$, Eq. (\ref{SquareK}) provides
$$  \square\,K^{A_1 A_2} \eq -18\,\Lambda\,K^{A_1 A_2} \ma  K^{BE}\,\Psi^{A_1 A_2}_{\phantom{A_1 A_2}BE}  \quad\textrm{ and }\quad
\square\,K_{A_1 A_2} \eq -18\,\Lambda\,K_{A_1 A_2} \ma  K_{BE}\, \Psi^{BE}_{\phantom{BE}A_1 A_2} \,.$$
The interesting thing happens when we have $p\geq 3$ in Eq. (\ref{SquareK}). For such a case, we obtain
\begin{align}
  \square\,K^{A_1\cdots A_p} \eq -(12+3p)\,\Lambda\,K^{A_1\cdots A_p} \ma (p-1) \,K^{BE(A_2\cdots A_{p-1}} \, \Psi^{A_p)A_1}_{\phantom{A_p)A_1}BE} \nonumber \,\,, \\
 \label{SquareK2} \\
 \square\,K_{A_1\cdots A_p} \eq -(12+3p)\,\Lambda\,K_{A_1\cdots A_p} \ma (p-1) \,K_{BE(A_2\cdots A_{p-1}} \, \Psi^{BE}_{\phantom{BE}A_p)A_1} \nonumber\,\,.
\end{align}
Then, contracting these equations with $\varepsilon_{ C D A_1 A_2}$ and $\varepsilon^{ C D A_1 A_2}$ respectively, we find the following integrability conditions
\begin{equation}\label{IntegrabDirac-p3}
  \varepsilon_{C D A_1 A_2 } K^{B E A_2 (A_3\cdots A_{p-1}} \, \Psi^{A_p)A_1}_{\phantom{A_p)A_1}BE} \eq 0 \quad\textrm{ and }\quad  \varepsilon^{C D A_1 A_2 } K_{B E A_2 (A_3\cdots A_{p-1}} \, \Psi^{BE}_{\phantom{BE}A_p)A_1} \eq 0 \,.
\end{equation}
These relations imply that the Weyl tensor must be algebraically special in order for the manifold to admit solutions of Eq. (\ref{ConfInv1}) when $p\geq 3$.

\subsection{Twistor Equation}

Now, we shall analyse Eq. (\ref{ConfInv2}) in the case $p=1$, namely the twistor equation. Consider a twistor of positive chirality $\psi^A$, \textit{i.e.} assume that
$$  \nabla_{AB}\,\psi^C \eq \zeta_{[A}\, \delta_{B]}^{\;C} \,. $$
Then, taking the covariant derivative of this equation and then inserting into the relation that defines the curvature of the spinorial connection, Eq. (\ref{SpinorCurvature2}), lead to
\begin{align}
  4\,\mathfrak{R}^{G\ph{F}C}_{\ph{A}F\ph{C}D}\, \psi^D \eq &  \varepsilon^{GABE}\,\left( \nabla_{AB}\,\zeta_{[E}\, \delta_{F]}^{\;C} \me  \nabla_{EF}\,\zeta_{[A}\, \delta_{B]}^{\;C} \right)  \nonumber \\
  \eq & -\,2 \, \nabla^{GC}\,\zeta_{F} \ma 2\,  \nabla^{GA}\,\zeta_{A} \, \delta_{F}^{\;C} \me   \nabla^{CA}\,\zeta_{A} \, \delta_{F}^{\;G} \,\,. \label{Integrab.Twistor1}
\end{align}
Taking the symmetric and the skew-symmetric parts of the above equation in the pair of indices $GC$, and also contracting the indices $C$ and $F$, lead eventually to the following relations:
\begin{equation}\label{Integ.Twistor2}
  \Psi^{GC}_{\phantom{GC}FD}\,\psi^D \eq 0 \quad,\quad \nabla^{GC}\,\zeta_F \eq -\, \frac{1}{2}\, \Phi^{GC}_{\phantom{GC}FD}\,\psi^D \me 2\,\Lambda\,\psi^{[G}\,\delta_{F}^{\;C]} \quad,\quad  \nabla^{CG}\,\zeta_C \eq  3\, \Lambda\,\psi^G \,.
\end{equation}
The first of these relations states that the Weyl tensor must be algebraically special in order for the manifold to admit a twistor. The analogous of this constraint in four-dimensional spacetimes is that the null vector associated to the twistor is a repeated principal null direction. The second relation in (\ref{Integ.Twistor2}) implies that in an Einstein manifold, namely when $\Phi^{GC}_{\phantom{GC}FD}$ vanishes, the spinor $\zeta_A$ is a twistor of negative chirality, as previously recognized in \cite{Kerr6D}. Finally, the third relation in (\ref{Integ.Twistor2}) guarantees that if the Ricci scalar vanishes, $\Lambda=0$, then the spinor $\zeta_A$ obeys the massless Dirac equation. Analogously,  if $\psi_A$ is a twistor of negative chirality, namely
$$ \nabla^{AB}\,\psi_C \eq \chi^{[A}\, \delta^{\;B]}_{C}  \, $$
then the following integrability conditions hold
$$  \Psi^{GF}_{\phantom{GF}DE}\,\psi_F \eq 0 \quad,\quad \nabla_{DE}\,\chi^G \eq -\, \frac{1}{2}\, \Phi^{GF}_{\phantom{GF}DE}\,\psi_F \me 2\,\Lambda\,\psi_{[D}\,\delta_{E]}^{\;G} \quad,\quad  \nabla_{CD}\,\chi^C \eq 3\, \Lambda\,\psi_D \,. $$
Therefore, analogous conclusions hold for twistors of negative chirality.

\subsection{Integrable Pure Spinor}

Finally, we shall consider the integrability condition for Eq. (\ref{ConfInvL}) when $p=1$, \textit{i.e.} we are going to analyse the integrability conditions for the existence of an integrable pure spinor. Let $\xi^A$ be a spinor of positive chirality such that its associated maximally isotropic distribution is integrable, in other words, Eq. (\ref{IntegrablePure+}) holds. Then, defining $\theta_B=\xi^A V_{AB}$, it follows that
\begin{equation}\label{Intreg.Pure2+}
  \xi^A\,\nabla_{AB} \,\xi^C \eq \theta_B \, \xi^C\,.
\end{equation}
In particular, note that $\xi^A\theta_A$ vanishes. Before proceeding, it is useful to define the differential operator $D_B\equiv \xi^A\nabla_{AB}$, in terms of which Eq. (\ref{Intreg.Pure2+}) is conveniently written as $D_{B}\xi^C=\theta_B \xi^C$. Differentiating this equation, one concludes that
\begin{equation}\label{Dtheta}
  D_{[A}\,D_{B]}\, \xi^C \eq  \xi^C \, D_{[A}\,\theta_{B]} \,.
\end{equation}
On the other hand, we have that
\begin{align*}
  \left( D_A\,D_B \me D_B\,D_A \right) \xi^C \eq & \, \xi^E \, \xi^F \left( \nabla_{FA}\,\nabla_{EB} \me  \nabla_{EB}\,\nabla_{FA} \right)\xi^C  \,.
   \end{align*}
Contracting the latter equation with $\varepsilon^{ABKL}$ and performing some algebraic manipulations, one eventually concludes that
$$ \varepsilon^{AB KL}\,D_A\,D_B\,\xi^C \eq  2\, \xi^D\,\xi^E\xi^{[K} \,\Psi^{L]C}_{\phantom{D]E}DE} \,. $$
Then, inserting Eq. (\ref{Dtheta}) in the left hand side of the latter relation yields
\begin{equation}\label{Integ.Cond.Pure}
   \xi^D\,\xi^E\xi^{[K} \,\Psi^{L]C}_{\phantom{D]E}DE} \eq \frac{1}{2}\, \xi^C\,  \varepsilon^{AB KL} \, D_{A}\,\theta_{B} \,.
\end{equation}
In particular, this integrability condition implies that the spinorial representation of the Weyl tensor must obey the following algebraic constraint:
\begin{equation}\label{Integ.Cond.Pure2}
 \xi^D\,\xi^E\xi^{[K} \,\Psi^{L][C}_{\phantom{D]E}DE}\,\xi^{A]} \eq 0 \,.
\end{equation}
Since the null structure defined $\xi^A$ has a projective nature, it is invariant by changes of scale, $\xi^A\rightarrow \hat{\xi}^A=e^{-\lambda} \xi^A$, where $\lambda$ stands for an arbitrary scalar function. Thus, a particularly interesting question is whether it is possible to choose the scale $\lambda$ in such a way that instead of (\ref{Intreg.Pure2+}) the simpler equation
\begin{equation}\label{Intreg.Pure3+}
  \hat{\xi}^A\,\nabla_{AB} \,\hat{\xi}^C \eq 0
\end{equation}
holds. Imposing (\ref{Intreg.Pure3+}) and using (\ref{Intreg.Pure2+}) one concludes that the function $\lambda$ must be such that $D_B \lambda = \theta_B$. However, this equation is a constraint over the spinorial field $\theta_B$ that generally admits no solution. Indeed, applying $D_A$ to the latter relation, and using the fact that the Levi-Civita connection is torsionfree, one concludes that $\theta_B$ must be such that $D_{[A}\theta_{B]}$ vanishes. Such condition, in turn, due to (\ref{Integ.Cond.Pure}), requires that
\begin{equation}\label{Integ-Dtheta}
  \xi^D\,\xi^E\xi^{[K} \,\Psi^{L]C}_{\phantom{D]E}DE}\,\zeta_C \eq 0\,,
\end{equation}
where $\zeta_C$ is any spinor such that $\xi^A \zeta_A\neq0$. Note that this condition is not guaranteed by the integrability condition (\ref{Integ.Cond.Pure2}). Obviously, completely analogous relations hold for the case of integrable pure spinors of negative chirality.

It is worth stressing that the integrability conditions presented here are necessary but, generally, not sufficient. Indeed, in order to obtain the sufficient conditions, one must further workout the integrability conditions obtained by manipulating higher derivatives of the analysed differential equation. This process should be continued until the integrability condition of order $n$ is a consequence of the integrability conditions of order less than $n$.  Where by an integrability condition of order $n$ it is meant the constraints obtained over the curvature by differentiating the analysed differential equation $n$ times. Here, we have presented just the first order integrability conditions. For instance, the integrability condition (\ref{Integ.Cond.Pure2}), for the pure spinor $\xi^A$ to generate an integrable maximally isotropic distribution, is not sufficient. Indeed, in Ref. \cite{HigherGSisotropic2} it has been obtained the sufficient condition for a maximally isotropic distribution to be integrable in any dimension, using the tensorial formalism, and in \cite{Bat-Spin6D} this condition has been translated to the six-dimensional spinorial language. The final result is that a maximally isotropic distribution generated by $\xi^A$ is integrable if, and only if,
$$  \xi^E\xi^{[K} \,\Psi^{L][C}_{\phantom{D]E}DE}\,\xi^{A]} \eq 0 \,,$$
which is stronger than the constraint (\ref{Integ.Cond.Pure2}). In particular, this constraint along with the traceless property of the spinorial representation of the Weyl tensor, $\Psi^{AB}_{\phantom{AB}AC}=0$, implies that Eq. (\ref{Integ-Dtheta}) holds. Nevertheless, it is worth recalling that Eq. (\ref{Integ-Dtheta}) is just the first order integrability condition for the equation $D_B \lambda = \theta_B$.

As a final comment, note that all integrability conditions presented here impose constraints on the part of the curvature involving only the Weyl tensor.  This should be expected from the fact that the equations investigated here are conformally invariant along with the fact that Weyl tensor is the only part of the curvature that is invariant under an arbitrary conformal transformation. The notion of algebraically special Weyl tensors have played an important role in four dimensions. For instance, it motivated the discovery of the Kerr solution \cite{Kerr} and, more generally, it has enabled a full analytic integration of Einstein's equation in vacuum spacetimes of Petrov type $D$ \cite{Kinnersly-typeD}. More recently, algebraic classifications for the Weyl tensor have also been defined in higher dimensions and used, for instance, in the study of spacetimes with vanishing curvature invariants, see
\cite{Rev. Ortaggio,5D class.,CMPP,Bat-art4} and references therein. Hopefully, the algebraic integrability conditions presented here will also be of relevance in this context.


\section{Conclusions and Perspectives}\label{Sec.Conclusions}

In this article the theme of conformal transformations and conformally invariant equations in the six-dimensional spinorial formalism have been exploited and many identities have apparently been introduced for the first time, as exemplifies (\ref{Identity6D}), (\ref{Conf_Connection1}), (\ref{CurvatureTransf}) and (\ref{HarmonicFields}). The use of the index notation has enabled the generalization of three well-known conformally invariant spinorial equations to conformally invariant equations for totally symmetric spinorial objects of arbitrary rank, \textit{i.e.} for fields of spin greater than 1/2. We have also managed to give some geometrical interpretation for such fields. Finally, the integrability condition for some of these equations have been worked out. In addition, the relevance of harmonic 3-forms in six dimensions have been pointed out. Particularly, closed (anti-)self-dual 3-forms can be seen as instanton solutions for the Yang-Mills theory on a gerbe.

In four dimensions, one important application of the spinorial formalism to conformal transformations is the study of asymptotically flat spacetimes \cite{Peeling-Penrose}. In particular, one interesting result is the so-called peeling theorem, which states that in asymptotically flat spacetimes the algebraic type of the Weyl tensor and, more generally, of a massless field, become more and more special as the null infinity is approached along a null geodesic \cite{Peeling-Penrose}. Then, a natural question is whether similar results can be proved in six dimensions using the tools introduced here. In reference \cite{Peel-Pravd.} the peeling theorem for the Weyl tensor was generalized to all dimensions using simple arguments about scaling properties. But this result was based on an assumption that latter has proved to be incompatible with a physically reasonable definition of asymptotically flat spacetime in dimensions greater than four. Just recently, the peeling theorem for the Weyl tensor has been correctly extended to higher-dimensional space-times, by Godazgar and Reall \cite{Peel-Reall}. In the latter reference it has been shown that there are qualitative differences between the four-dimensional and the higher-dimensional cases. These differences stem from the fact that in higher dimensions it is more involved to impose the basic physical requirements that an asymptotically flat spacetime should satisfy, namely it must be stable under small perturbations, should admit the existence of gravitational radiation, and should be compatible with the existence of a generator for the Bondi energy \cite{Hollands-Asymp.Flat, Asymp.Flat_Tanabi}. Although the peeling property for the Weyl tensor have already been extended to higher dimensions, the study of the asymptotic behaviour  of a general massless field in an asymptotically flat spacetime remains to be done. It is fair to expect that the tools introduced here can be valuable for the analysis of this problem in six dimensions.

It is worth stressing that the power of choosing a suitable notation cannot be underestimated. Indeed, the choice of using the index notation was one of the main reasons for the numerous triumphs accomplished by Roger Penrose on the application of the spinorial formalism in four-dimensional general relativity. Analogously, the present article hints that similar achievements can be performed in six dimensions. Curiously, it turns out that there are great resemblances between the spinorial formalism in four and six dimensions. For instance, a four-component Weyl spinor in six-dimensional spacetimes can be conveniently written as a two-component object over the quaternion field \cite{Kugo,Bengtsson}, which stems from the isomorphism $SPin(\mathbb{R}^{1,5})\sim SL(2,\mathbb{H})$. Thus, a spinor in six dimensions can be written in a fashion quite similar to Penrose's index notation, using indices ranging from 1 to 2. This fact can lead to nice insights on generalizations of four-dimensional results to six dimensions. In particular, it would be interesting to investigate whether the relation between complex conformal transformations and torsion, pointed out by Penrose in four dimensions \cite{Penrose-Torsion}, is carried to six dimensions.

Further investigation in the direction of this article is already in progress. The aim of the forthcoming work is to use the index notation of the six-dimensional spinorial formalism to study conformal Killing-Yano tensors and their connection with Killing spinors. In particular, we shall work out the integrability conditions for the latter objects. Conformal Killing-Yano tensors have acquired great relevance in the study of higher-dimensional black holes, as they allowed the analytical integration of several field equations as well as the geodesic equation on the Kerr-NUT-(A)dS spacetimes in arbitrary dimensions, see \cite{Frolov_KY} and references therein. Moreover, Killing spinors are of central importance in supergravity and general supersymmetric theories \cite{Cariglia}.

\section*{Acknowledgments}
I want to thank some colleagues in the Physics Department of Universidade Federal de Pernambuco for the warm reception. I also would like to thank the anonymous referees for the careful analysis and for some valuable suggestions.


\end{document}